\documentclass[twocolumn,nofootinbib]{revtex4}
\usepackage{graphicx}
\usepackage{float}
\usepackage{amsmath,amssymb,amsfonts}
\usepackage{bbm}

\newcommand{\bse}{\begin{subequations}}
\newcommand{\ese}{\end{subequations}}
\newcommand{\be}{\begin{equation}}
\newcommand{\ee}{\end{equation}}
\newcommand{\bea}{\begin{eqnarray}}
\newcommand{\eea}{\end{eqnarray}}
\newcommand{\ba}{\begin{array}}
\newcommand{\ea}{\end{array}}

\input amssym.def
\input amssym.tex

\usepackage[colorlinks=true, linkcolor=blue, bookmarks=true, citecolor=red]{hyperref}

\begin{document}

\title{From Schwinger Balls to Curved Space}

\author{Davood Allahbakhshi\footnote{allahbakhshi@ipm.ir}}
\affiliation{School of Particles and Accelerators, Institute for Research in Fundamental Sciences (IPM),
P.O.Box 19395-5531, Tehran, Iran}

\begin{abstract}
%\abstract{
It is shown that the Reissner-Nordstrom black hole is also a gravitational Schwinger ball. It is also shown that both massless and massive-particle gravitational Schwinger balls are thermodynamic systems by deriving the first law of thermodynamics for them. Inconsistency between classical geometrical and microscopic definitions of the horizon is discussed. We propose a new metric, more consistent with microscopic picture of black hole, as gravitational Schwinger ball, by speculations. It has some interesting features.
%}
\end{abstract}

\maketitle

%\tableofcontents
\section{Introduction}
Hawking radiation \cite{Hawking:1974sw} showed a conceptual conflict between the classical geometrical definition of black holes and local quantum field theories. Thermal evaporation of the black hole with no memory of the initial state of the system raised the problem of information paradox \cite{Hawking:1976ra}. Bekenstein entropy \cite{Bekenstein:1973ur} with no explanation in classical picture is another problem. Many attempts, to solve these problems with no real success, have raised serious doubts about the correctness and physical acceptability of the classical definition of black holes itself \cite{Hawking:2014tga}. It seems that this classical picture is not completely physical and there should be a new definition of black holes with more physical characteristics.

Recent \emph{graviton ball} picture of black holes \cite{Dvali-Gomez-papers} has shed light on problems of these mysterious objects. Particularly the \emph{Schwinger ball} picture can produce and explain many different characteristics of Schwarzschild black hole as well as problems like information paradox \cite{Allahbakhshi:2016ctk}. Bekenstein entropy, black hole thermodynamics, the relation between Schwarzchild radius and the mass of the black hole can be derived from this picture. Also it is shown that the Hawking radiation is not consistent with Schwarzschild state except for eternal black hole \cite{Allahbakhshi:2016dwq}.

If black holes are realy Schwinger balls, we should be able to explain other types of black holes than Schwarzschild black hole in this picture. The first one in the list is Reissner-Nordstrom black hole. In present paper we show that the Reissner-Nordstrom black hole is also a gravitational Schwinger ball. The only point is including the energy of the electromagnetic field appropriately. At the same time this explanation reveals an inconsistency between the microscopic and classical macroscopic definitions of the black hole. While in general relativity the black hole is the space-time behind the horizon as a coordinate singularity, in the microscopic picture it is a region of significant Schwinger effect. Favouring the microscopic Schwinger ball picture leads us to a new definition of the classical geometry which carries facts about the quantum theory of gravity.

\section{The Form of A Black Hole Metric}
The relation between the size and the mass of the Schwarzschild black hole is
\be
R^{d-3}=\frac{16\pi GM}{c^2(d-2)\Omega_{d-2}}
\ee
and in previous work \cite{Allahbakhshi:2016ctk}, we found that the number of gravitons inside the Schwinger ball (black hole) is
\be
N=\frac{d-2}{8\pi}\;\frac{A}{L_P^{d-2}},
\ee
where $A$ is the area of the ball, as is also disscused in \cite{Dvali-Gomez-papers}, which states that the relation between the entropy and the number of gravitons of a Graviton ball is $S=2\pi N/(d-2)$. Two equations above determine the microscopic state of the Schwarzschild black hole. But what about charged black holes? There is a complicated relation between the size of the charged black hole and its mass and charge
\be
R_\pm^{d-3}=G\mu \left( 1 \pm \sqrt{1 - \frac{q^2}{\mu^2}} \right),
\ee
where
\bea
\mu &&= \frac{8\pi M}{c^2(d-2)\Omega_{d-2}}\cr\cr
q^2 &&= \frac{8\pi Q^2}{c^4\varepsilon _0(d-3)(d-2)\Omega ^2_{d-2}}.
\eea
How can we derive it from Schwinger effect, like what we did for Schwarzschild black hole \cite{Allahbakhshi:2016ctk}? Although the answer to this question is so simple, it leads us to a new world of physical concepts.

The Schwarzschild radius above is the root of the metric function of the black hole
\be
f(R)=1-\frac{2G\mu}{R^{d-3}}+\frac{Gq^2}{R^{2(d-3)}}.
\ee
The question is this: How can the root of the metric fuction be related to Schwinger effect?

The answer to the question above is interesting! Let us start by Schwarzschild black hole. The metric function is
\be
f(R)=1-\frac{2G\mu}{R^{d-3}}.
\ee
We know that the root of this function is where the Schwinger effect becomes significat. In fact we can rewrite the function above in this form
\be\label{metric-function}
f(R) = 1-\frac{a(R)}{a_g(R)},
\ee
where 
\be
a(R) = \frac{G\mu c^2}{R^{d-2}}
\ee
is the Newtonian acceleration (or surface gravity in this case) at radius $R$ and
\be
a_g(R)=\frac{c^2}{2R}
\ee
is the acceleration, needed to create a graviton which can live in a sphere of radius $R$, by Schwinger effect\footnote{Remember that the acceleration needed to create a quanta of mass $m$ by gravitational Schwinger effect is $m c^3/\hbar$ \cite{Allahbakhshi:2016ctk}.}. It means that the metric function is a criterion which determines that \emph{how much we are away from Schwinger limit.}

Also for Reissner-Nordstrom black hole we can write the metric function in the form of \ref{metric-function}, but this time we have
\be\label{Newtonian-acceleration}
a(R) = \frac{G\mu _*(R)c^2}{R^{d-2}},
\ee
where
\bea
\mu _*(R) &&= \frac{8\pi M_*(R)}{c^2(d-2)\Omega_{d-2}}\cr\cr
M_*(R) &&= M - \frac{Q^2}{2c^2(d-3)\varepsilon_0 \Omega_{d-2}R^{d-3}}.
\eea
$M_*(R)$ is the mass/energy inside the sphere of radius $R$, since from elementary physics we know that just the energy inside this region contributes to the gravitational force but not the mass/energy outside it. So we need to subtract the electromagnetic energy, outside the sphere, from the total mass/energy. To see this we note that $M_*$ can be written in this form
\be\label{mstar}
M_* = M_b + \frac{E(R , 0)}{c^2} = M - \frac{E(\infty , R)}{c^2},
\ee
where
\be
E(x,y)=- \frac{1}{2\varepsilon_0 \Omega_{d-2}}\int _x ^y{\frac{Q^2}{r^{d-2}}\;dr}
\ee
is the energy of the spherically symmetric electromagnetic field between radii $x$ and $y$. $M_b$ is the \emph{bare mass} of the black hole which absorbs the infinite electromagnetic energy in $E(R,0)$ to produce the \emph{physical mass} $M$. From \ref{mstar} we simply conclude that
\be
M = M_b + \frac{E(\infty , 0)}{c^2},
\ee
as expected.

Note that the acceleration \ref{Newtonian-acceleration} is the \emph{Newtonian acceleration} at radius $R$, not the \emph{surface gravity}!

\section{The First Law of Thermodynamics}
The size of a gravitational Schwinger ball which creates particles of mass $m$ can be derived by equiating the \emph{Newtonian acceleration} to \emph{Schwinger limit} $a_c=mc^3/\hbar$
\be
\frac{mc^3}{\hbar} = \frac{G\mu _*(R)c^2}{R^{d-2}},
\ee
or in terms of $M_*$ it is
\be
\frac{mc^3}{\hbar} = \frac{8\pi GM _*(R)}{(d-2)\Omega_{d-2}R^{d-2}}.
\ee
But $A=\Omega_{d-2}R^{d-2}$ is the area of the Schwinger ball, so we have
\be\label{master-first-law}
\frac{(d-2)c^3}{8\pi G\hbar}\;mA = M _*(R).
\ee
Varying this relation gives the first law of black hole thermodynamics. In next two parts we find this law for a graviton Schwinger ball and a massive-particle gravitational Schwinger ball.

\subsection{Graviton Schwinger Ball}
The variation of the equation \ref{master-first-law} for a graviton of mass $m=\hbar/(2cR)$ is
\be\label{master-var}
\frac{(d-2)c^3}{8\pi G\hbar}\;(dm\;A+m\;dA) = dM _*(R).
\ee
After some simple algebra and using the relations
\bea
dm &&= -\frac{\hbar}{2cR^2}dR\cr\cr
dA &&= (d-2)\Omega_{d-2}R^{d-3}dR.
\eea
It is easy to see that
\be
dE = \frac{c^2\kappa}{2\pi}\;\frac{dA}{4G} + V dQ,
\ee
where $E=Mc^2$ is the energy of the Schwinger ball, and
\be
\kappa = \frac{(d-3)c^2}{2}\left[ \frac{1}{R} - \frac{q^2}{R^{2d-5}} \right]
\ee
is nothing but the surface gravity at the horizon and
\be
V(R)=\frac{Q}{(d-3)\varepsilon_0 \Omega_{d-2} R^{d-3}}
\ee
is the electric potential at the surface of the Schwinger ball.

\subsection{Massive-Particle Schwinger Ball}
For massive-particle Schwinger ball, which creates massive particles instead of gravitons \footnote{In fact such Schwinger balls also produce gravitons and the Schwinger radius of gravitons (massless particles) is generally much larger than the Schwinger radius of massive particles . So the massive-particle Schwinger balls are hidden inside the graviton Schwinger balls. To become more familiar with massive-particle Schwinger balls see \cite{Allahbakhshi:2016ctk}.}, the calculation is exactly like we did in previous part but this time since the mass of the massive particle is fixed, $dm = 0$ in equation \ref{master-var} and we have
\be
dE = \frac{c^2\kappa}{2\pi}\;\frac{dA}{4G} + V dQ,
\ee
with
\be
\kappa = \frac{c^2}{\lambda} + \frac{(d-3)c^2}{2}\left[ \frac{2}{\lambda} - \frac{q^2}{R^{2d-5}} \right],
\ee
where $\lambda$ is the \emph{Compton wavelength} of the particle of mass $m$, and $R$ can be derived from equation \ref{master-first-law}. So massive-particle Schwinger balls also obey the first law of thermodynamics but with different temperature.

So we see that the Reissner-Nordstrom black hole is also a \emph{gravitational Schwinger ball}, but this time we should include the energy of the electromagnetic field. The very important question here is that why the metric function is of the form of \ref{metric-function}? And why does it have a root at the Schwinger limit? In the next section we discuss these questions.

\section{About Curved Space}
As mentioned, the metric function of Reissner-Nordstrom and so Schwarzschild metrics can be written in the form 
\be
f(r)=1-\frac{a(r)}{a_g(r)}.
\ee
It means that the criterion for defining the \emph{effective time or length} intervals is the Schwinger limit. Consider \emph{proper} time-like and space-like intervals in a black hole geometry
\bea
dT &&= \sqrt{1-a/a_g}\;dt\cr\cr
dX &&=\frac{dx}{\sqrt{1-a/a_g}}.
\eea
These expressions are very similar to time-dilation and length-contraction in \emph{special relativity}, but this time contraction and dilation coefficients are made of $\sqrt{1-a/a_g}$ instead of $\sqrt{1-v^2/c^2}$. It means that the acceleration is responsible for contraction and dilation of the spatial and temporal intervals of (effective and emergent curved) space-time and the reference acceleration is the \emph{Schwinger limit}!

On the other hand, the metric function is zero at the horizon and makes a coordinate singularity, from the viewpoint of the static observer. In microscopic picture, the horizon is where the Schwinger effect becomes significant but, in geometrical picture it is where the space ends! On the other hand the Schwinger limit is a \emph{smooth limit} at which the probability for a particle to be produced, is larger than $e^{-1}$ after that, not a definite point at which something special happens sharply!

From the Schwinger effect viewpoint, the coordinate singularity at the horizon, that the metric is \emph{time-dependent} beyond it, can be considered as the tool that classical effective theory, general relativity, uses to produce particles. This sigularity is similar to Hagedorn temperature in statistical bootstrap model of hadrons. In Hagedorn's model, the Hagedorn temperature is the largest achievable temperature by the hadron gas. But simultaneusly it shows signs of a phase transition. Later we found that at Hagedorn temperature the hadron gas experiences a phase transition to deconfined matter. But the Hagedorn's model can not explain beyond this temperature because is not capable to do it. Also we found that at low baryonic chemical potentials this transition is in fact a smooth transition, a crossover. In general relativity, as a classical effective theory, the horizon is the deepest achievable place in the space, by static observer outside the black hole but, from microscopic point of view it is where a smooth transition occures from where the Schwinger effect is not significant to where it is. In other words the microscopic theory of gravitons does not need a horizon to produce gravitons through Schwinger effect, since in this picture the Schwinger effect is nothing but decays of gravitons to gravitons (or other particles).

The discussion above shows that, from microscopic point of view, where we have included the \emph{quantum effects of gravitons}, the horizon is not a special point but a smooth region and we need a smooth metric there, even from the viewpoint of the static observer. Such smooth non-singular metric should be derived emergently from a quantum theory of gravity, but here we propose a metric just by speculations. Since the weight of the Schwinger effect is $e^{-a_g/a}$, one may propose a metric of the form
\be
ds^2=-f(r)\;dt^2+\frac{dr^2}{f(r)}+r^2\;d\Omega_{d-2}^2,
\ee
with the metric function
\be\label{proposed-metric-function}
f(r)=e^{-a(r)/a_g(r)}.
\ee
When we are far from the Schwinger limit (horizon) and so $a \ll a_g$, the function can be expanded
\be
f(r)\approx 1- \frac{a(r)}{a_g(r)},
\ee
which is nothing but the function \ref{metric-function}. Note that the function \ref{proposed-metric-function} can produce the same results of \ref{metric-function}, in any practical case that we have tested the general relativity. Here we have considered $g_{rr}=1/g_{tt}$, but more generally $g_{rr}$ can be $e^{\chi}/g_{tt}$ in which $\chi$ is another function. This metric obviously is not a solution of the pure Einstein-Hilbert action but another unknown one which should be found.

\begin{figure}
\includegraphics[scale=.5]{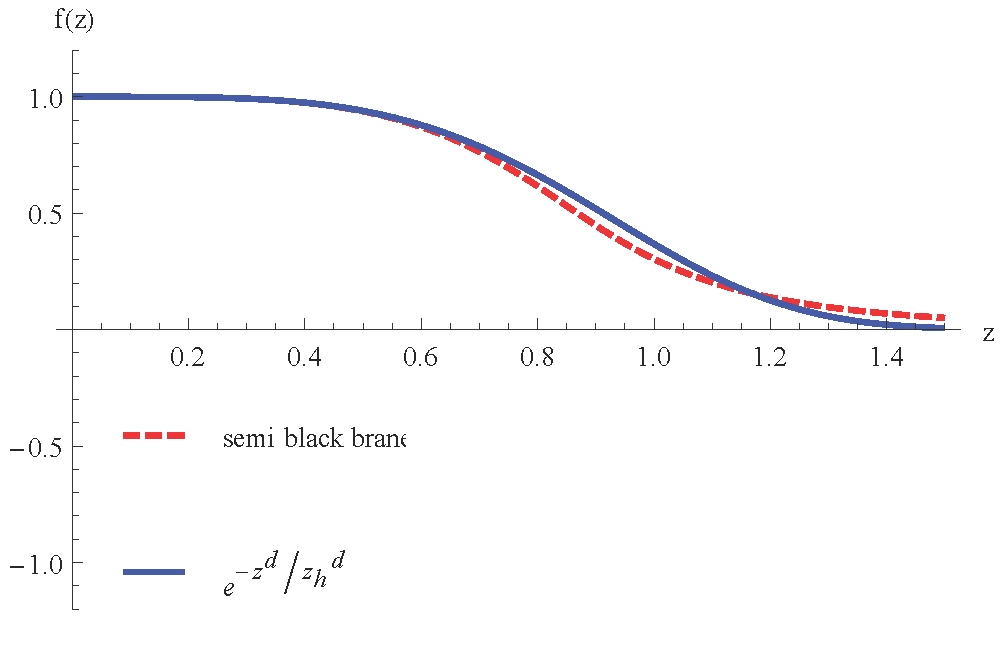}
\caption{\label{comp}The metric function of semi black brane and our proposed fuction.}
\end{figure}

We do not have any more evidence for correctness of the metric above but, there is just an interesting similarity between this metric and one we found in \cite{Allahbakhshi:2016gyj}. For this to be more clear we note that for an AdS black brane, similar to the metric above, we can again propose a metric of the form
\be
ds^2=\frac{1}{z^2}\left(-f(z)\;dt^2+\frac{dr^2}{f(z)}+d\vec{x}_{d-1}^2\right),
\ee
with
\be
f(z)=e^{-z^d/z_h^d}.
\ee
In figure \ref{comp} we have plotted the function above and what we found for \emph{semi black brane} in \cite{Allahbakhshi:2016gyj}.

For justifying our proposed metric we can claim that including the effect of quantum fluctuations of gravitons about a classical geometry in a \emph{mean field approximation} may lead to an effective action similar to what we found in \cite{Allahbakhshi:2016gyj}. Such effective action generally can include the Schwinger effect and may have a \emph{semi black hole} solution, like what we proposed above. This claim of course needs investigations.

\section{Summary and Discussion}
We saw that the Reissner-Nordstrom metric can also be described as a gravitational Schwinger ball, but we need to include the energy of the electromagnetic field appropriately. One may ask why does the Reissner-Nordstrom metric just include \emph{gravitational Schwinger effect} but not \emph{electric Schwinger effect}? The answer is obvious. This metric is the solution of the Einstein-Maxwell action which does not include the quantum effects of QED. If we use an electromagnetic effective action like Heisenberg-Euler action, in addition to the \emph{effective interaction terms} between gravity and electromagnetism, then its solution should include both gravitational and electric Schwinger effects.

Inspired by Schwinger weight, we also proposed a metric, consistent with microscopic picture of black holes as Schwinger balls. This metric has interesting features. First of all it produces usual gravitational effects. Secondly it is similar to semi black brane that we found in our previous work. It also has another interesting feature; If we calculate the surface gravity of this metric we find
\be\label{Yukawa-like}
\kappa = \frac{GM}{r^{d-2}} \; e^{-2GM/r^{d-3}},
\ee
which can be written in the form
\be
\kappa = \frac{GM}{r^{d-2}}e^{-m(r)\;r},
\ee
with
\be\label{dynamical-mass}
m(r)=\frac{2GM}{r^{d-2}}.
\ee
$m(r)$ can be considered as the \emph{dynamical running mass} of the graviton, produced by the background mean field at radius $r$. In fact it is the energy of the graviton which can be produced through Schwinger effect at radius $r$ by \emph{Newtonian acceleration}. The surface gravity \ref{Yukawa-like} is a Yukawa like force produced by a massive particle. But there is a very important difference. The effect of the mass in ordinary Yukawa theories is important just at IR but not UV, while the gravitational acceleration \ref{Yukawa-like} is Newtonian at IR limit but, completely different at UV. In fact the proposed dynamical mass of the graviton \ref{dynamical-mass} is \emph{infinite} at UV, which makes the theory \emph{Renormalizable}. On the other hand the acceleration \ref{Yukawa-like} is zero at $r\rightarrow 0$ limit, for $d>3$. It means that the hypothetical quantum theory of gravity, which produces such geometry in classical effective limit, is \emph{asymptotically free}.

If our speculations are nearly correct then we can hope that the simple quantized general relativity is non-renormalizable because its UV behaviour is not correct, since it is probably a \emph{truncated effective action}.

More investigations of course give more interesting results.

%%%%%%%%%%%%%%%%%%%%%%%%%%%%%%%%%%%%%%%%%%%%%%%%%%%%%%

\end{document}